\documentclass[12pt,reqno]{amsart}
\pdfoutput=1 
\usepackage{amsmath,bbm}
\usepackage{latexsym}
\usepackage{amsfonts}
\usepackage{amssymb}
\usepackage{color}
\usepackage{graphicx}
\usepackage{url}
\usepackage{enumerate}
\usepackage{tikz}
\usetikzlibrary{shadings, intersections, calc, plotmarks}
\usepackage{marginnote}
\usepackage{stackrel}

\usepackage{geometry}
\xdefinecolor{tumblue}     {RGB}{0,101,189}
\xdefinecolor{tumgreen}    {RGB}{162,173,  0}
\xdefinecolor{tumred}      {RGB}{229, 52, 24}
\xdefinecolor{tumivory}    {RGB}{218,215,203}
\xdefinecolor{tumorange}   {RGB}{227,114, 34}
\xdefinecolor{tumlightblue}{RGB}{152,198,234}

\newtheorem{theorem}{Theorem}
\newtheorem*{theorem*}{Theorem}
\newtheorem{lemma}{Lemma}
\newtheorem{corollary}{Corollary}
\newtheorem*{corollary*}{Corollary}

\newtheorem{definition}{Definition}

\newcommand{\comment}[1]{\blue{\footnote{\blue{#1}}}} 
\renewcommand{\comment}[1]{} 

\newcommand{\note}[1]{} 

\newcommand{\blue}[1]{\textcolor{blue}{#1}}

\newcommand{\id}{{\rm{id}}} 
\newcommand{\spec}{{\rm{spec}}}

\newcommand{\CP}{{\mathbb{C}}{\bf P}}
\newcommand{\cH}{\mathcal{H}}
\newcommand{\R}{\mathbbm{R}}

\newcommand{\C}{\mathbbm{C}}
\newcommand{\N}{\mathbbm{N}}
\newcommand{\cL}{\mathcal{L}}

\newcommand{\cT}{\mathcal{T}}
\newcommand{\cC}{\mathcal{C}}

\newcommand{\cG}{\mathcal{G}}

\newcommand{\1}{\mathbbm{1}}

\def\>{{\rangle}}
\def\<{{\langle}}

\newcommand{\be}{\begin{equation}}
\newcommand{\ee}{\end{equation}}
\newcommand{\bea}{\begin{eqnarray}}
\newcommand{\eea}{\end{eqnarray}}

\newcommand{\tr}[1]{\mathrm{tr}\left[#1\right]} 

\newcommand{\margintxt}[1]{} 

\begin{document}

\title[Q. tomography from the evolution of a single expectation]{ Quantum tomography from the evolution of a single expectation}

\author[Rall]{Hjalmar Rall$^{1,2}$}
\email{hjalmar.rall@tum.de}
\author[Wolf]{Michael M. Wolf$^{1,2}$}
\email{m.wolf@tum.de}
\address{$^1$ Department of Mathematics, Technical University of Munich}
\address{$^2$ Munich Center for Quantum
Science and Technology (MCQST),  M\"unchen, Germany}

\begin{abstract} We investigate the possibility of performing full quantum tomography based on the homogeneous time evolution of a single expectation value. Remarkably, every non-trivial binary measurement evolved by any quantum channel, except for a null set, in principle enables full quantum state tomography. We show that this remains true when restricted to Lindblad semigroups, although unitary evolution---even with added simply depolarizing noise---is insufficient beyond the qubit case, highlighting the necessity of non-trivial noise. We establish an analog of Takens' embedding theorem for quantum channels, which  incorporates prior information into the framework. We also provide estimation bounds for finite statistics, and analyze the feasibility of recovering an infinite time series of expectation values from a finite one using only spectral properties of the evolution. 
\end{abstract}

\maketitle
\tableofcontents

\section{Introduction}\label{sec:intro}

Every first course in quantum theory teaches us how to compute the evolution of an expectation value when we are given a dynamical equation (like Schr\"odinger's or Heisenberg's) together with a mathematical description of (i) the state and (ii) the observable. 

In this paper, we turn the perspective around and ask whether (i) or (ii) can be inferred from the evolution of an expectation value that is for instance given as a finite time series. In other words, we ask how much information is contained in the time evolution of an expectation value or the evolution of the probability of a single measurement outcome. We investigate this question for finite-dimensional quantum systems: if a known observable/state undergoes a known homogeneous time evolution, what can be learned about the state/observable of the system from the time series? To this end, we first explore in Sec.\ref{sec:nogo}, when and how an infinite time series can be recovered from a finite part of it without invoking any knowledge about the underlying state or observable---only using spectral information about the evolution. This exposes limitations of the information content of the time series and allows us to prove some no-go theorems for full quantum tomography in Sec.\ref{sec:ngresults}.  In particular,  beyond the tomography of single-qubit states, any unitary-based (and in this sense \emph{noise-free}) evolution of an expectation value or probability is always insufficient for full quantum tomography and remains so when adding simply depolarizing noise. 

In contrast, we will see in Sec.\ref{sec:result} that almost every noisy evolution turns out to enable full quantum tomography of either state or observable if a sufficiently large part of the time series is known. The number of required points is then related in the Takens'-inspired Thm.\ref{thm:TakensQ} to the Minkowski-dimension of the unknown set of states or observables. 

While up to this point, we assume that the expectation values are known precisely, we will analyze aspects of finite statistics in Sec.\ref{sec:stability}. 

Finally, in Sec.\ref{sec:qubits} we shed more light on the case of two-dimensional Hilbert spaces. This turns out to be a special case in the sense that the no-go result for unitary-based full quantum tomography holds for observables but not for states. 
\vspace*{5pt}

\paragraph{\bf Related work: } Our work has three roots: \emph{Takens' delay embedding theorem} from dynamical systems theory \cite{Takens81,Sauer1991}, the concept of \emph{observability} in control theory \cite{KALMAN1960491,skelton1988dynamic} and, of course, \emph{quantum tomography}---especially under prior information \cite{Heinosaari2013,Carmeli_2016}.

Takens' delay embedding theorem is an adaption of \emph{Whitney's embedding theorem} from differential topology. In essence, Taken's theorem is a non-linear version of Thm.\ref{thm:main1d2} and Thm.\ref{thm:TakensQ} below. However, the non-linear version does not immediately imply our linear results: while the set of considered maps is restricted, the allowed perturbations are also constrained. Therefore, terms such as  `generic' or `almost all' are not easily comparable between the linear and non-linear frameworks. In fact, we see in the example of unitary evolutions that a na\"{\i}ve adoption of Takens' result fails to hold in the linear case. Moreover, different from our results, Takens' theorem is formulated for a fixed evolution and almost all (non-linear) measurements. Nevertheless, we make extensive use of the ideas developed in the vicinity or aftermath \cite{Sauer1991,robinson_2010} of Takens' theorem. 

In the language of control theory, Takens' non-linear result and our linear counterparts may be summarized as `observability is generic'. This is, in fact, a long-known result (see \cite{genericobs,genericobs78}) but, as far as we know, it has not been worked out with the details presented here in the framework of quantum evolutions.

In quantum information theory, `delayed embeddings' were used in \cite{AssessingD} to assess the dimension of a quantum system and in \cite{Mendl} for unitary process tomography.

Quantum tomography, in general, is a broad topic with a wealth of results, many of which focus on statistical efficiency \cite{Flammia_2012,10.1145/2897518.2897544,Elben2023}. It should be clear that the scheme discussed in the present paper does not aim to offer a competitive alternative, but rather explores the limit at which quantum tomography is possible at all. 

There are a number of papers, some of which we became aware of only after completion of this work, that address the type of questions we investigate: \emph{to what extent can dynamics enhance the possibility of quantum tomography?}

Most notably, it was shown in \cite{PRAFlammia2010} that repeated application of a discrete unitary evolution is insufficient for the reconstruction of quantum states in dimensions $d\geq3$. In \cite{Czerwinski2021} it was observed that, specifically for $d\leq 3$, random unitary dynamics can enhance the feasibility of quantum tomography. Moreover, it has been shown in \cite{Kech16} for specific invertible channels that semialgebraic sets of states can be recovered from the evolution of a small number of  POVM elements, and a similar, albeit less explicit, discussion of this question has been addressed in \cite{JAMIOKOWSKI2000469} for Lindbladian semigroups.

In a similar spirit, but using a slightly different framework that allows for time-dependent Hamiltonians, 
\cite{PhysRevLett.124.010405} shows that
for fully controllable quantum systems subject to a random field,  the evolved expectation value of any observable almost always enables full quantum state tomography. 

Finally, concurrent with our work, \cite{peruzzo2025reconstructingquantumstateslocal} independently showed that randomized parameter choices for Lindblad-type evolutions yield observability with probability one.


\section{Preliminaries}\label{sec:prelim}
We denote by $\cH_d$ the set of hermitian matrices in $\C^{d\times d}$ and by $\cH_d^{(0)}\subset\cH_d$ and $\cH_d^{(1)}\subset\cH_d$ the subspace and subset of matrices that have trace zero and trace one, respectively.  $\cH_d$ is a $d^2$-dimensional real vector space and it will occasionally be useful to make this fact explicit by invoking a vector space isomorphism $\nu:\cH_d\rightarrow\R^{d^2}$. Via $\nu$ the Lebesgue measure on $\R^{d^2}$ defines a measure on $\cH_d$. Assuming that $\nu$ maps a Hilbert-Schmidt orthogonal basis with first element $\1$ onto the standard-basis in $\R^{d^2}$, every hermiticity preserving, unital linear map $T:\cH_d\rightarrow\cH_d$ is represented by a transformation matrix $\hat{T}:=\nu\circ T\circ\nu^{-1}$, whose first column is given by $(1,0,\ldots,0)$. We denote the set of all such maps $T$ by $\cT_d$. 
The dual\footnote{Here, \emph{duality} is understood w.r.t. the Hilbert-Schmidt inner product. That is, $T^*$ is such that $\tr{XT(Y)}=\tr{T^*(X)Y}$ for all $X,Y$.} map $T^*$ of any $T\in\cT_d$  is then trace-preserving and Hermiticity-preserving. Moreover, with $Q:\cH_d\rightarrow\cH_d^{(0)}$, $Q(X):=X-\1\tr{X}/d$  the orthogonal projector onto the trace-less subspace, every $T\in\cT_d$ satisfies $QT=QTQ$. Applying this identity to the spectral equation, we see that $QT$, understood as endomorphism on $\cH_d^{(0)}$, satisfies  $\spec(QT)\cup\{1\}=\spec(T)$.

As an affine space
$\cT_d$ is isomorphic to $\R^{d^2\times (d^2-1)}$. 
We will call a subset of $\cT_d$ a \emph{null set} if the corresponding subset in $\R^{d^2\times (d^2-1)}$ has zero Lebesgue measure. In this respect, the subset $\cC_d\subset\cT_d$ of completely positive maps, i.e., the set of \emph{quantum channels} represented in the Heisenberg picture, has non-zero measure. This is most easily seen by noting that the completely depolarizing channel $H\mapsto\1\tr{H}/d$, which has eigenvalues $(1,0,\ldots,0)$,  lies in the interior of $\cC_d$, because its Choi matrix does. 
There are other ways of defining a measure on  $\cC_d$, but all natural choices of measures are known to be equivalent (cf. \cite{RandomChannels}).

An important subset of quantum channels consists of those that are elements of quantum dynamical semigroups. We denote this set by
 \begin{subequations}
\begin{eqnarray}
    \cL_d&:=&\big\{T\in\cT_d \;\big|\;\exists L\in\cG_d: T=e^L \},\quad\text{where}\\
    \cG_d&:=& \big\{L:\cH_d\rightarrow \cH_d\;\big|\; \forall t\in[0,\infty): e^{tL}\in\cC_d\big\},
\end{eqnarray}
 \end{subequations}
is the set of \emph{Lindblad generators}. Note that, since $L(\1)=0$ for every $L\in\cG_d$, the corresponding transformation matrix $\hat{L}:=\nu\circ L\circ \nu^{-1}\in \R^{d^2\times d^2}$ has first column $(0,\ldots,0)$. Consequently, $\cG_d$ is a subset of a $d^2(d^2-1)$ dimensional real vector space. The following shows that within this space,  $\cG_d$ has non-zero measure.
\begin{lemma}\label{lem:Ld} The set $\cG_d$ has non-zero $d^2(d^2-1)$-dimensional Lebesgue measure. Similarly, the set $\cL_d$ has non-zero measure in $\cT_d$ while its boundary $\partial\cL_d$ is a null set.
\end{lemma}
\begin{proof}
    We will show that subsets of $\cG_d$ and $\cL_d$ are smooth manifolds of  dimension $d^2(d^2-1)=\dim(\cT_d)$. To this end, we use that every $L\in\cG_d$ is \emph{conditionally completely positive} \cite{EvansCCP}. More precisely, there is a completely positive map $\Phi$ and $K\in\C^{d\times d}$ such that  $L(X)=\Phi(X)-KX-XK^*$ and $\Phi(\1)=K+K^*$. That is, the hermitian part of $K$ is fully determined by $\Phi$, while the anti-hermitian part of $K$ is unconstrained. With $|\Omega\rangle:=\sum_{i=1}^d|i,i\rangle$ we denote by $C\in\C^{d^2\times d^2}$, $C:=(L\otimes\id)(|\Omega\rangle\langle\Omega|)$ the Choi matrix of $L$. Let $(B_i)_{i=1}^{d^2}$ be any Hilbert-Schmidt orthonormal operator basis of $\cH_d$ with $B_{d^2}=\1/\sqrt{d}$. Then $|b_i\rangle:=(B_i\otimes\1)|\Omega\rangle$ is an orthonormal basis of $\C^{d}\otimes\C^d$. Expressed in this basis, the Choi matrix corresponds to an admissible generator $L$ iff it is of the form 
    \begin{equation}\label{eq:ChoiL}
        C= \begin{pmatrix} P & v
    \\ v^* & -\tr{P}
        \end{pmatrix}, 
    \end{equation}
    where $P\in\cH_{d^2-1}$ is positive-semidefinite and $v\in\C^{d^2-1}$ is such that its real part is determined by $P$ (as it corresponds to $K+K^*$) and its imaginary part is unconstrained.\footnote{In fact, if we assume that $\Phi$ has block diagonal Choi matrix $P\oplus 0$, then $v_i=-\sqrt{d}\; \tr{B_i K}$ for all $i<d^2$.} In this way, Eq.(\ref{eq:ChoiL}) together with the Choi isomorphism gives rise to an injective linear parametrization $\R^{d^2(d^2-1)}\ni\hat{C}\mapsto C\mapsto L$. Furthermore, if $P$ is positive definite, a sufficiently small open neighborhood of an admissible $\hat{C}$ will be mapped entirely into $\cG_g$. This neighborhood is then a smooth manifold of dimension $(d^2-1)^2+(d^2-1)=d^2(d^2-1)$, which proves the first claim of the Lemma. 
    
    This manifold is then mapped further into $\cT_d$ by the matrix exponential. As this is again smooth and locally injective, also $\cL_d$ contains a smooth manifold whose dimension coincides with the one of the ambient space. Hence, $\cL_d$ has non-zero measure in $\cT_d$. Moreover, as the map from $C$ in Eq.(\ref{eq:ChoiL}) to $e^L$ is a local diffeomorphism and therefore maps null sets to null sets and only boundaries to boundaries, $\partial\cL_d$ is a null set since the boundary of positive semidefinite matrices is.
\end{proof}

Bearing in mind that both $\cL_d$ and $\cC_d$ have non-zero measure in $\cT_d$, the following shows that almost all quantum dynamical semigroups and quantum channels have a non-degenerate spectrum\footnote{By this we mean that all eigenvalues have algebraic multiplicity equal to one.}. This seems to be  `well-known', but since we could not find an explicit proof, we provide one for completeness:\footnote{It has been brought to our attention that a different proof has been provided recently in \cite{diagonalizable24}.}
\begin{lemma}\label{lem:deg}
    The subset of maps in $\cT_d$ that have degenerate spectrum is a null set.
\end{lemma}
\begin{proof}
    Representing $T\in\cT_d$ by a real transformation matrix $\hat{T}$, whose first column is $(1,0,\cdots,0)$, we see that the proof can be boiled down to showing that in $\R^{(d^2-1)\times(d^2-1)}$ the subset of degenerate matrices is a null set. A square matrix $M$ is degenerate iff the discriminant of its characteristic polynomial vanishes. Expressing the discriminant in terms of the resultant of the characteristic polynomial and its derivative, we can regard the discriminant $q$  itself as a polynomial function whose variables are the $(d^2-1)^2$ entries of $M$. Hence, the set of degenerate matrices forms a real algebraic variety specified by $q(M)=0$. However, such zero sets are null sets \cite{Mityagin2020} unless $q$ is constant zero, which is clearly not the case since the set of non-degenerate matrices is not empty.
\end{proof}
In order to be able to work with constrained sets that do not necessarily have any further structure (i.e., no manifolds), we use the following notion of dimension, which is always well-defined:

\begin{definition}[Covering number \& Minkowski dimension]
For any bounded subset $S$  of a finite-dimensional normed space, one defines:
\begin{itemize}
\item 
The \emph{covering number} $N(\epsilon,S):=\min\big\{m\in\N\;|\; S\subseteq\bigcup_{i=1}^m B_\epsilon (x_i),\; x_i\in S\big\}$, where $B_\epsilon (x_i)$ denotes the open ball of radius $\epsilon>0$ around $x_i$.
\item The \emph{Minkowski dimension} (a.k.a. \emph{upper box counting dimension})\\ $D(S):=\limsup_{\epsilon\rightarrow 0}\frac{\ln N(\epsilon,S)}{\ln(1/\epsilon)}$. 
\end{itemize}
\end{definition}

To understand the definition of the Minkowski dimension, it is helpful to observe that if $S$ is a $d$-dimensional manifold, then $N(\epsilon,S)$ scales as $ (1/\epsilon)^d$. The Minkowski dimension utilizes this to define the dimension for arbitrary sets while preserving the familiar notion of dimension where it is already defined otherwise. In fact, if $S$ is a $d$-dimensional manifold, then $D(S)=d$. For instance, the Minkowski dimension of the manifold of density matrices of rank $r$ in $\cH_d$ is $2dr-r^2-1$ \cite{Grabowski_2005}.  Other properties of the Minkowski dimension are that it is (i)  non-increasing under Lipschitz continuous maps, (ii) non-increasing under taking subsets and (iii) such that for $S-S:=\{x-y|x,y\in S\}$ we have $D(S-S)\leq 2D(S)$ (cf.\cite{robinson_2010}, chap.3).

\section{Series extensions}\label{sec:nogo}

In this section, we will show that a small number of at most $d^2$ data points of an observed time series is sufficient to reconstruct the entire series with the help of the spectrum of the time evolution map. The simpler this spectrum, the smaller the number of required data points. On the flip side of this observation are no-go results for tomography: if all the information of the entire time series is already contained in a few data points, the series may not contain enough information for a tomographic reconstruction of the underlying quantum state (or observable). We begin with the discrete-time case.

For any vector space endomorphism, and in particular, for any quantum channel $T\in\cC_d$, we define $\delta(T)$ to be the degree of its minimal polynomial, i.e., the minimal degree of any non-zero polynomial $p$ for which $p(T)=0$. The Cayley-Hamilton theorem implies that $\delta(T)\leq d^2$. Moreover, we define $j_0(T)$ to be the size of the largest Jordan block of $T$ that corresponds to a zero eigenvalue, and $j_0(T):=0$ if $T$ has a trivial kernel ($\ker{(T)}=\{0\}$).

\begin{figure}[t]
    \centering
    \includegraphics[width=0.9\textwidth]{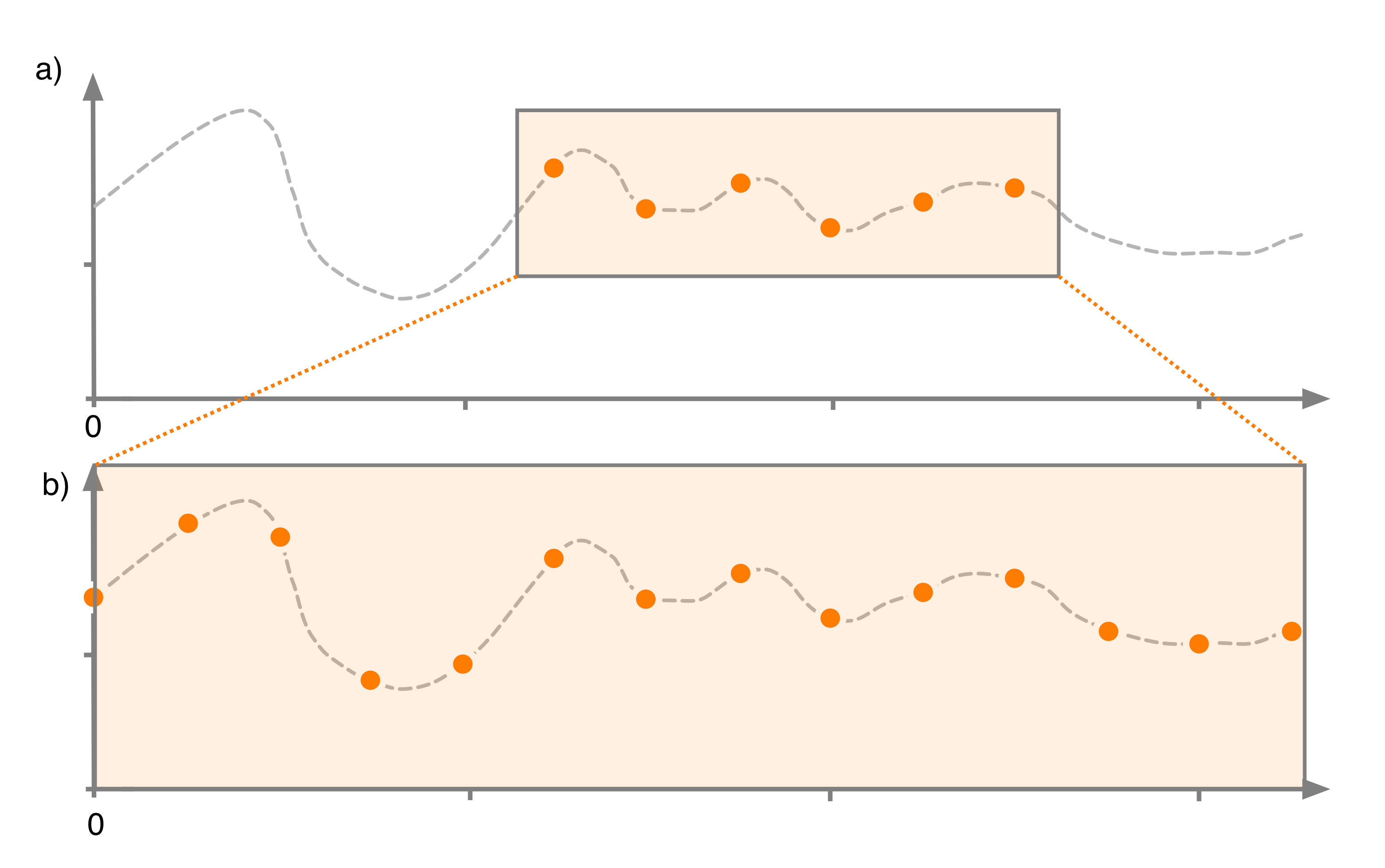} 
    \caption{Time-series extension. Suppose a sufficient number of consecutive data points of a homogeneous discrete-time evolution of a $d$-dimensional quantum system w.r.t. a known quantum channel is given. Then the series can be extended to its infinite counterpart by a linear map that does neither depend on the state, nor on the observable (Thm.\ref{thm:seriesextension}). Similar applies to continuous time evolution: generically, $d^2$ arbitrarily spaced data points suffice to complete the graph without knowing state or observable (Thm.\ref{thm:Lextension}).} \label{fig:1}
\end{figure}

\begin{theorem}\label{thm:seriesextension}
    Consider sequences $a=(a_i)_{i=0}^\infty\in\R^\infty$ of the form $a_i:=\tr{\rho T^i(H_0)}$ with $T^0:=\id$.
    \begin{enumerate}
        \item For every $t_0\in\N_0$, every  $T\in\cT_d$   and any $t\geq\delta(T)-\kappa$ with $\kappa:=\min\{t_0,j_0(T)\}$, there is a linear map $E:\R^{t}\rightarrow\R^\infty$ such that for all $H_0,\rho\in\cH_d$: 
         \begin{equation}
             E:(a_i)_{i=t_0}^{t+t_0-1} \ \mapsto\ (a_i)_{i=\kappa}^\infty.
         \end{equation}
         \item For every $T\in\cT_d$ there is a subspace $V\subset \R^\infty$ of dimension $\dim(V)\leq \delta(T)\leq d^2$ such that for all $H_0, \rho\in\cH_d$: $a\in V$.
    \end{enumerate}

\end{theorem}
\begin{proof} \emph{(1):}
    The minimal polynomial of $T$ is of the form $p(x)=\prod_{\lambda}(x-\lambda)^{j_\lambda}$ where the product runs over all distinct eigenvalues $\lambda$ of $T$ and $j_\lambda$ is the size of the largest Jordan block of $T$ corresponding to the eigenvalue $\lambda$ (cf. \cite{horn13}, Thm. 3.3.6). Since the degree of monomials in $p$ thus ranges from $j_0$ to $\delta:=\delta(T)$,  the equation $p(T)=0$  implies that we can express both $T^{j_0}$ and $T^\delta$ as a linear combination of monomials of higher and lower degree, respectively. More generally, we have 
    \begin{eqnarray}\label{eq:Oe7g3f1a}
        T^{t_0-k} &=&\sum_{i=0}^{\delta-j_0-1} b_i \; T^{t_0+i-k+1},\quad\text{for all }k\in\{1,\ldots, t_0-j_0\},\\
        T^{\delta+k} &=& \sum_{i=0}^{\delta-j_0-1} c_i\;  T^{j_0+i+k},\quad\text{for all }k\in\N_0,\label{eq:Oe7g3f1b}
    \end{eqnarray}
    with suitable constants $b_i,c_i$, which can be determined from the minimal polynomial of $T$. By applying the maps in Eqs.(\ref{eq:Oe7g3f1a},\ref{eq:Oe7g3f1b}) to $H_0$ and taking the trace with $\rho$, we obtain the same equations with  $T^i$ replaced by $a_i$. Hence, if $t_0>j_0$,  Eq.(\ref{eq:Oe7g3f1a})  allows us to extend the time series, which is initially of length $t$ starting at $t_0$, step-by-step to smaller times until we reach $j_0$. Due to the linear combinations, every single step in this extension corresponds to a linear map ($\R^t\rightarrow\R^{t+1}$, $\R^{t+1}\rightarrow\R^{t+2}$, etc.) so that their composition is again linear.
    In a similar vein, Eq.(\ref{eq:Oe7g3f1b}) allows an inductive extension of the time series to larger times by means of consecutive linear maps. Note that both cases require a `seed' of length $t\geq\delta-j_0$ and monomials of sufficiently large degree, which amounts to  $t+t_0\geq \delta$. Combining these inequalities leads to $t\geq\delta-\kappa$.\vspace*{5pt}

    \emph{(2)} follows from \emph{(1)} by realizing that $a$ can be obtained from $(a_0,\ldots,a_{\delta-1})\in\R^\delta$ by a linear map $E:\R^\delta\rightarrow\R^\infty$ that does not depend on $\rho$ or $H_0$. Hence, the range $E(\R^\delta)=:V$ fulfills the claim.
\end{proof}

That map that extends a finite to an infinite time series in Thm.\ref{thm:seriesextension} depends only on (the spectrum of) $T$. If we allow for an additional dependence on $H_0$, then we can further reduce the number of required data points by one:
\begin{theorem}\label{thm:seriesextension2}
    For $T\in\cT_d$ define $T_Q:\cH_d^{(0)}\rightarrow\cH_d^{(0)}$, $T_Q:=Q\circ T$. Consider 
      sequences $a=(a_i)_{i=0}^\infty\in\R^\infty$ of the form $a_i:=\tr{\rho T^i(H_0)}$ with $T^0:=\id$.
    \begin{enumerate}
        \item For every $(T,H_0)\in\cT_d\times\cH_d$, every $t_0\in\N_0$   and any $t\geq\delta(T_Q)-\kappa$ with $\kappa:=\min\{t_0,j_0(T_Q)\}$, there is an affine map $A:\R^{t}\rightarrow\R^\infty$ such that for all $\rho\in\cH_d^{(1)}$: 
         \begin{equation}
             A:(a_i)_{i=t_0}^{t+t_0-1} \ \mapsto\ (a_i)_{i=\kappa}^\infty.
         \end{equation}
         \item For every $(T,H_0)\in\cT_d\times\cH_d$ there is an affine subspace $V\subset \R^\infty$ of dimension $\dim(V)\leq \delta(T_Q)\leq d^2-1$ such that for all $ \rho\in\cH_d^{(1)}$: $a\in V$.
    \end{enumerate}
\end{theorem}
\begin{proof}
    The main difference from the previous theorem is that now we restrict $\rho$ to the \emph{affine} space $\cH_d^{(1)}$ instead of working with the linear space $\cH_d$. In order to transfer the argument, we will first translate $\cH_d^{(1)}$ to $\cH_d^{(0)}$ by subtracting any $\rho_0\in\cH_d^{(1)}$, then work with the linear space $\cH_d^{(0)}$ and finally undo the translation and add the offset again. This is displayed by the identity
    \begin{eqnarray}
       a_i\ = \ \tr{\rho T^i(\cH_0)} &=& \tr{\rho_0 T^i(H_0)}+\tr{(\rho-\rho_0) T^i(H_0)}\nonumber\\
        &=& \tr{\rho_0 T^i(H_0)}+\tr{(\rho-\rho_0)T_Q^i (H_Q)},\label{eq:QTQai}
    \end{eqnarray}
    where $H_Q:=Q(H_0)$ and we have used that $(\rho-\rho_0)=Q(\rho-\rho_0)\in\cH_d^{(0)}$ and $Q\circ T= Q\circ T\circ Q$. For $T_Q^i:=(T_Q)^i$ that appears in the last term of Eq.(\ref{eq:QTQai}), we can now argue exactly as in Eqs.(\ref{eq:Oe7g3f1a},\ref{eq:Oe7g3f1b}) and express monomials with exponent $i$ outside $I:=[t_0,t+t_0-1]\subset\N_0$ as a linear combination of monomials that lie inside this range (unless $i<\kappa$). Then we can exploit Eq.(\ref{eq:QTQai}) again in order to express everything in terms of the original sequence $(a_i)$ plus an offset $\gamma=\gamma(T,H_0)$ that does not depend on $\rho$. That is, with suitable coefficients $B_{ik}$ that depend solely on the spectrum of $T_Q$ (which in turn is determined by the spectrum of $T$) we obtain:
    \begin{eqnarray}
        a_i &=& \tr{\rho_0 T^i(H_0)} +\sum_{k\in I} B_{ik}\;\tr{(\rho-\rho_0) T_Q^k(H_Q)}\nonumber\\
        &=& \tr{\rho_0 T^i(H_0)} +\sum_{k\in I} B_{ik}\big(a_k-\tr{\rho_0 T^k(H_0)}\big)\nonumber\\
        &=:& \gamma_i + \sum_{k\in I} B_{ik}\; a_k.
    \end{eqnarray}
    From here, the two statements of the theorem follow in the same way as for the previous theorem. The bound $\delta(T_Q)\leq d^2-1$ is due to the fact that $\cH_d^{(0)}$, on which $T_Q$ acts, is $(d^2-1)$-dimensional.
\end{proof}

Next, we consider the continuous-time case and show that, generically, $d^2$ data points, arbitrarily located in time, are again sufficient to construct a unique extension to the entire real line.  

\begin{theorem}\label{thm:Lextension}\margintxt{Is there a simple way to reduce the number of required points by one again as in Thm.\ref{thm:seriesextension2}?}
    Let $t_1,\ldots, t_{d^2}\in[0,\infty)$ be distinct and consider a function of the form $a(t):=\tr{\rho e^{tL}(H_0)}$. For every $L\in\cG_d$ except for a null set, the following is true:
    
    There are analytic functions $\beta_1,\ldots,\beta_{d^2}$ depending only on $(t_k)_{k=1}^{d^2}$ and on the spectrum of $L$ s.t. for any $\rho, H_0\in\cH_d:$
    \begin{equation}
        a(t)=\sum_{k=1}^{d^2} \beta_k(t) a(t_k)\qquad\text{for all } t\in\R.
    \end{equation}
    This holds in particular for every $L$ with non-degenerate, real spectrum.
\end{theorem}
\begin{proof}
Assume that $L$ has distinct eigenvalues $\lambda_1,\ldots,\lambda_{d^2}$. According to Lem. \ref{lem:deg} this assumption only excludes a null set.
    The Cayley-Hamilton theorem allows us to express $e^{tL}$ as a polynomial in $L$. More precisely,
    \begin{equation}\label{eq:CHexp1}
        e^{tL}=\sum_{i=0}^{d^2-1} y_i(t) L^i,\quad\text{with}\quad y_i(t)=\sum_{j=1}^{d^2} C_{i,j} e^{\lambda_j t},
    \end{equation}
    where $C$ is the inverse of a confluent Vandermonde matrix depending on the $\lambda_j$'s (see \cite{LutherRostExp,HedemannExp}). Our goal is to find functions $\beta_k(t)$ such that 
    \begin{equation}
        e^{tL}=\sum_{k=1}^{d^2} \beta_k(t) e^{t_k L}= \sum_{i=0}^{d^2-1} \underbrace{\left(\sum_{k=1}^{d^2} \beta_k(t) y_i(t_k)\right)}_{=y_i(t)} L^i,
    \end{equation}
    where the second equality uses Eq.(\ref{eq:CHexp1}) for every $e^{t_k L}$. The sought functions can then be obtained as solutions of the linear equation $Y \beta(t)=z(t)$ with $Y_{j,k}:=e^{\lambda_j t_k}$ and $z_j(t):=e^{\lambda_j t}$. This solution exists if $Y$ is invertible. If the $\lambda_j$'s are real, $Y$ is a well-studied case of a generalized Vandermonde matrix, which is known to be always invertible (see \cite{YANG2001201} or Sec. 4.2. in\cite{pinkus_2009}). If complex eigenvalues appear, this is no longer guaranteed. However, since in the absence of degeneracy eigenvalues depend analytically on the matrix entries, the map $L\mapsto \det(Y)$ is real analytic when we regard $L$ as an element of a $d^2(d^2-1)$ dimensional real vector space, parameterized using Eq.(\ref{eq:ChoiL}). Since this map is not constant zero, its zero set has measure zero \cite{Mityagin2020}. Therefore, except for a null set in $\cG_d$, $Y$ is invertible, which in turn shows that the desired $\beta_k$'s exist.\margintxt{Maybe, we should provide the $\beta_k$'s explicitly (e.g. using \cite{HedemannExp}) in an appendix.}
\end{proof}

\section{No-go results}\label{sec:ngresults}

We now use the extension results of the previous section to prove no-go results for time-series based quantum tomography by changing the viewpoint: if a small part of a time series has a unique extension, then the corresponding infinite time series can be significantly compressed, and thus can only contain a limited amount of information.
\begin{corollary} \label{cor:nogo}Let $\rho_0,H_0\in\cH_d$, $S\subset\cH_d$, $T\in\cT_d$ and denote by $\delta$ the degree of the  minimal polynomial of $T$. Consider the maps 
\begin{eqnarray}\label{eq:Takensmapcor1}
    S\ni\rho\ &\mapsto& \big(\tr{\rho T^i(H_0)}\big)_{i=0}^\infty,\\
    S\ni H &\mapsto& \big(\tr{\rho_0 T^i(H)}\big)_{i=0}^\infty.\label{eq:Takensmapcor2}
\end{eqnarray}
\begin{enumerate}
    \item If $S$ is a topological manifold of dimension $D(S)>\delta$, then  neither of the maps in Eqs.(\ref{eq:Takensmapcor1},\ref{eq:Takensmapcor2}) is  injective.
    \item\margintxt{Can one strengthen this? E.g. to $D(S-S)>\delta$?} If $S$ is a set whose Minkowski dimension satisfies $D(S)>\delta$, then neither of the maps in Eqs.(\ref{eq:Takensmapcor1},\ref{eq:Takensmapcor2}) can have a Lipschitz continuous inverse.
\end{enumerate}
\end{corollary}
\begin{proof} We will lay out the proof for the map in Eq.(\ref{eq:Takensmapcor1})---the proof for the second map is completely analogous.
From Thm.\ref{thm:seriesextension} we know that it suffices to consider $S\ni\rho\mapsto \alpha(\rho):=(a_0,\ldots,a_{\delta -1})\in\R^{\delta}$ since the map in Eq.(\ref{eq:Takensmapcor1}) is a composition $E\circ\alpha$. 

\emph{(1)}:  $\alpha$ is continuous. If it were also injective, the map $\beta:S\rightarrow\R^\delta\times\R^{D(S)-\delta}, x\mapsto (\alpha(x),0)$  would be a continuous injection. Since $S$ is locally homeomorphic to $\R^{D(S)}$, this would contradict the \emph{invariance of domain theorem} (e.g.\cite{HatcherAT}, Thm.2B.3).

\emph{(2)} Suppose there were a Lipschitz continuous inverse. Then, by the basic properties of the Minkowski dimension (cf.\cite{robinson_2010}, chap.3), $\delta=D(\R^\delta)\geq D(\alpha(S))\geq D(\alpha^{-1}\circ\alpha(S))= D(S)$, which contradicts the assumption.     
\end{proof}
In order to exemplify this result, we bound the degree of the minimal polynomial of an important class of maps, which contains all unitary channels as well as the simply depolarizing ones:
\begin{lemma}\label{lem:deltabound}
    For any unitary $U\in U(d)$, density matrix $\sigma\in\cH_d$, and $\lambda\in[0,1]$, the quantum channel $T_\lambda(H):=(1-\lambda) U^* H U+\lambda \1\tr{H \sigma}$ has minimal polynomial of degree $\delta(T_\lambda)\leq d^2-d+2-\delta_{\lambda,0}$.\footnote{Here, $\delta_{\lambda,0}\in\{0,1\}$ is the Kronecker delta.} 
\end{lemma}
\begin{proof} We use the fact that a $k$-fold degeneracy of any eigenvalue of a diagonalizable endomorphism reduces the degree of its minimal polynomial by $k-1$ (implied by  Thm. 3.3.6 of \cite{horn13}).

    First, consider the case of a unitary channel for which $\lambda=0$. The transformation matrix $\hat{T_0}$ is, up to a similarity transformation, equal to $U\otimes\overline{U}$. Since the latter (and therefore $T_0$) is diagonalizable and has an eigenvalue $1$ with multiplicity at least $d$, we get $\delta(T_0)\leq d^2-(d-1)$ as claimed.

    Let $V\in U(d^2)$ of the form $V=1\oplus v$ with $v\in U(d^2-1)$ be a diagonalizing unitary of $\hat{T}_0$ for which $(V\hat{T}_0 V^*)_{ii}=1$ for all $i\leq d$. As the only non-zero elements of $\hat{T}_1$ are in its first row, with the first element equal to $1$, the same is true for $V\hat{T}_1V^*$ due to the block structure of $V$. Consequently, $V\hat{T}_\lambda V^*$ can have non-zero elements only in its first row and on its diagonal, which is of the form $(1,1-\lambda,\ldots,1-\lambda,\ldots)$. If $\lambda\neq 0$, then $1-\lambda$ is a $(d-1)$-fold degenerate eigenvalue, while the matrix is still diagonalizable (as a complete set of eigenvectors can readily be constructed). Hence, if $\lambda\neq 0$, then $\delta(T_\lambda)\leq d^2-(d-2)$.
\end{proof}
\emph{Remark:} It should be clear from the proof that the bound is tight as for a generic unitary $U\in U(d)$ the matrix $U\otimes\bar{U}$ has no further degeneracies.\vspace*{5pt} 

Combining Lem.\ref{lem:deltabound} and Cor.\ref{cor:nogo} leads to the following no-go results:

\begin{corollary}\label{cor:nogo2}
    Consider $d>2$, $H_0,\rho_0\in\cH_d$ and $T_\lambda$ a quantum channel as specified in Lem.\ref{lem:deltabound}. 
    \begin{enumerate}[(i)]
        \item For  any natural number $r$ for which $d-\sqrt{d-3+\delta_{\lambda,0}}<r\leq d$ and any set  $S\subseteq\cH_d$ that contains all density operators of rank $r$, the map $S\ni\rho\mapsto\big(\tr{\rho T_\lambda^i(H_0) }\big)_{i=0}^\infty$ is not injective.
        \item The map $\cH_d\ni H\mapsto\big(\tr{\rho_0 T_\lambda^i(H) }\big)_{i=0}^\infty$ is not injective.
    \end{enumerate}
     
\end{corollary}
\begin{proof} \emph{(i)}: From Lem.\ref{lem:deltabound} we get $\delta(T_\lambda)\leq d^2-d+2-\delta_{\lambda,0}$. Then $\delta(T_\lambda)<D(S)$ since $D(S)\geq 2dr-r^2-1$. So we can apply Cor.\ref{cor:nogo} \emph{(1)} and obtain that the time-series given by the evolution of the expectation value is insufficient to uniquely identify the states in $S$, irrespective of $U$ and $H_0$.

\emph{(ii)}: Again by Lem.\ref{lem:deltabound} we have $\delta(T_\lambda)\leq d^2-d+2$, which for $d>2$ is strictly smaller than $D(\cH_d)=d^2$ so that the result follows as before from Cor.\ref{cor:nogo}.
\end{proof}
In particular, for unitary (and in this sense noise-free) evolution, this implies that full tomography is impossible for observables if $d\geq 2$ and for states if $d\geq 3$.
The latter thus recovers the result of \cite{PRAFlammia2010}.

Note that the case $d=2$ is special in two ways: first, the dimension $d^2-1$ of the set of density operators equals the degree $d^2-d+1$ of the minimal polynomial of a generic unitary channel. Second, the dimension $d^2$ of the set of Hermitian operators equals the degree  $d^2-d+2$ of the minimal polynomial of a mixture of a generic unitary and a simply depolarizing channel.
Hence, the above no-go results do not apply. A detailed discussion of the $d=2$ case will be given in Sec.\ref{sec:qubits}.\vspace*{5pt}

For more general evolutions, Cor.\ref{cor:nogo} leaves open the possibility of complete time-series-based tomography of states or observables if only one of them is known (and this will be the content of the following sections). However, when both the state and the observable are unknown, the time series together with knowledge of the evolution map is not sufficient to identify both:
\begin{corollary}
    Let $T\in\cT_d$ and assume $S_1,S_2\subset\cH_d$ are manifolds of dimension $d_1$ and $d_2$, respectively. If $\delta(T)<d_1+d_2$ , then the following map is not injective:
    \begin{equation}\label{eq:iuzugf3q}
        S_1\times S_2\ni (\rho,H)\ \mapsto \ \Big(\tr{\rho T^i(H)}\Big)_{i=0}^\infty =:a
    \end{equation}
\end{corollary}
\begin{proof}
    This is a variant of Cor.\ref{cor:nogo}\;\emph{(1)}: $S_1\times S_2$ is a manifold of dimension $d_1+d_2$ while we know from Thm.\ref{thm:seriesextension}\;\emph{(2)} that $a$ lies in a vector space of dimension at most $\delta(T)$. So if the continuous map in Eq.(\ref{eq:iuzugf3q}) were injective, then $\delta(T)<d_1+d_2$ would contradict the invariance of domain theorem.
\end{proof}
Here are two examples, which use that $\delta(T)\leq d^2$:
\begin{enumerate}
    \item If $S_1\simeq\CP^{d-1}$ is the set of all pure states and $S_2=\cH_d$ the set of all observables, then $d_1=2d-2$ and $d_2=d^2$. Since $d_1+d_2>d^2$ there is no $T\in\cT_d$ s.t. the resulting time evolution would allow an identification of state and observable.
    \item If $S_1$ is the set of all density matrices and $S_2$ a two-parameter family of observables, then again $d_1+d_2=(d^2-1)+2>d^2$, which makes joint time-series-based tomography impossible.\margintxt{What if $S_2$ is a one-parameter family?}
\end{enumerate}

\begin{figure}[t]
    \centering
    \includegraphics[width=1\textwidth]{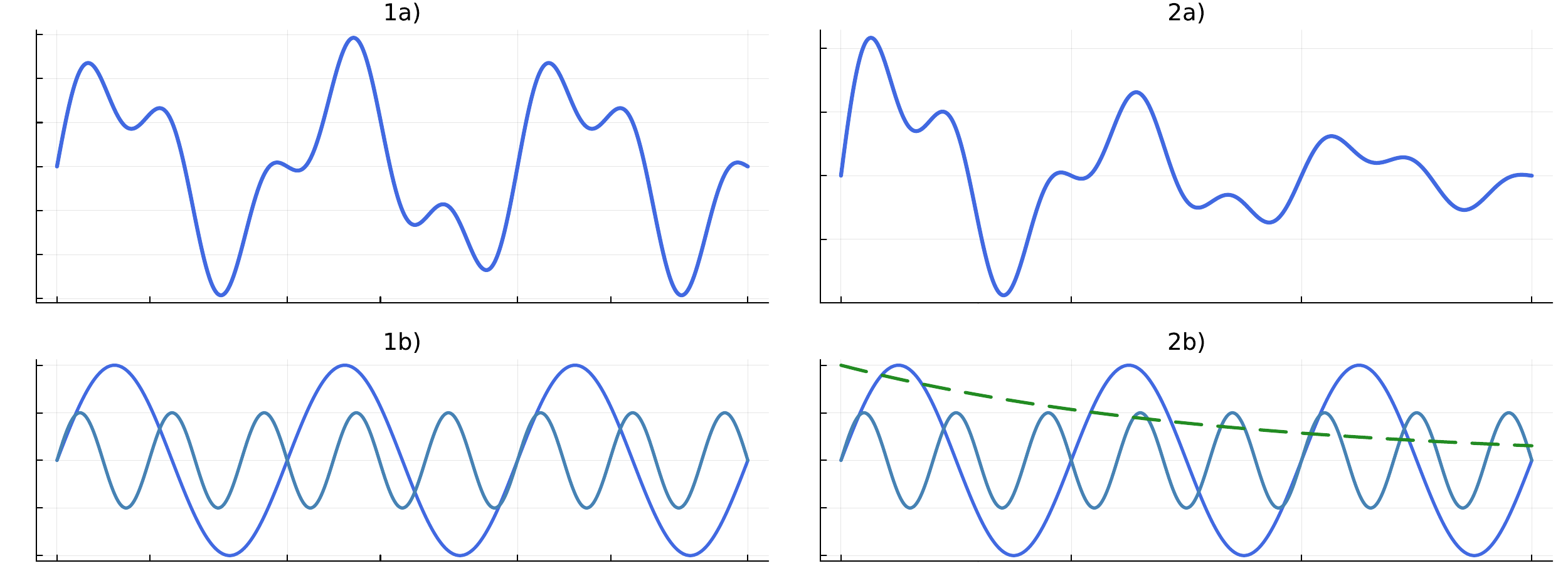} 
    \caption{Simplified graphical depiction of why noise added to unitary evolution can be beneficial for full quantum tomography from time evolution. \emph{Left}: Unitary/coherent evolution of an expectation value leads to a periodic signal (1a) whose information is carried in the frequency components (1b). \emph{Right}: If noise is added, here by a superimposed decay leading to the signal (2a), additional information is carried by the decay components (dashed line in 2b).} \label{fig:2}
\end{figure}

\section{Quantum  tomography}\label{sec:result}
We now turn to the positive results that show that, and under which conditions, time-series-based tomography of states (or observables) is possible. We begin with \emph{full tomography}, which is covered by Thm.\ref{thm:main1d2} and Cor.\ref{cor:contmain2} in the discrete and continuous setting, respectively. 

Before we get into the mathematical details, let us provide an informal but potentially more tangible explanation of why, on the one hand, unitary evolution is insufficient for complete quantum tomography and, on the other hand, why additional noise can help overcome this obstacle. Unitary evolution gives rise to a periodic signal whose information is carried in the coefficients of the frequency components (see Fig.\ref{fig:2}). However, as  spectral analysis reveals, there are at most $d^2-d+1$ different frequencies---too few when compared to the number of parameters of a general observable (or state). If sufficiently generic noise is added, the signal displays extra rates of decay, which potentially carry additional information. In fact, generically, this enables full tomography:    

\begin{theorem}\label{thm:main1d2} Let $t_0\in\N_0$ and $t_1,d\in\N$.
\begin{enumerate}
    \item For $t_1-t_0\geq d^2-2$ consider the  map $\alpha:\cH_d^{(1)}\rightarrow\R^{1+t_1-t_0} :$
    \begin{equation}
    \alpha(\rho):= \big(\tr{\rho T^i(H_0)}\big)_{i=t_0}^{t_1},\quad\text{with } T^0:=\id.
\end{equation}
\begin{enumerate}
    \item[(1.a)] For every $H_0\in\cH_d$ that is not proportional to the identity, the set of quantum channels $T\in\cC_d$ for which $\alpha$ is not injective is a null set. Moreover, the same is true when we replace $\cC_d$ by $\cL_d$.\margintxt{Part \emph{(1)} is also true when $d^2$ non-consecutive points in time are used. For part \emph{(2)} this is probably not true.}
    \item[(1.b)] For every quantum channel $T\in\cC_d$ with $d^2$ distinct eigenvalues, the set of $H_0$ for which $\alpha$ is not injective is a null set in $\cH_d$. 
\end{enumerate}\vspace*{5pt}
\item For $t_1-t_0\geq d^2-1$ consider the  map $\beta:\cH_d\rightarrow\R^{1+t_1-t_0} :$
    \begin{equation}
    \beta(H):= \big(\tr{\rho_0 T^i(H)}\big)_{i=t_0}^{t_1},\quad\text{with } T^0:=\id.
\end{equation}
\begin{enumerate}
    \item[(2.a)] For every $\rho_0\in\cH_d\setminus\cH_d^{(0)}$  the set of quantum channels $T\in\cC_d$ for which $\beta$ is not injective is a null set. Moreover, the same is true when we replace $\cC_d$ by $\cL_d$.
    \item[(2.b)] For every quantum channel $T\in\cC_d$ with $d^2$ distinct eigenvalues, the set of $\rho_0\in\cH_d^{(1)}$ for which $\beta$ is not injective is a null set in $\cH_d^{(1)}$. 
\end{enumerate}
\end{enumerate}

\end{theorem}
\emph{Remarks:} 1. Note that injectivity of the maps $\alpha,\beta$ implies automatically that their inverses are Lipschitz continuous as they are isomorphisms between finite-dimensional vector spaces. 2. If we regard the space of Hermitian matrices as equipped with the Hilbert-Schmidt norm, the minimal Lipschitz constant of the inverse maps is given by the inverse of their smallest singular value. 3. The map $\alpha$ becomes injective on all of $\cH_d$ under the same conditions if we add one more data point, i.e., demand that $t_1-t_0\geq d^2-1$.
\begin{proof}
     Like every boundary of a bounded, convex set, $\partial\cC_d$ is a null set, as is $\partial L_d$ by Lem.\ref{lem:Ld}. Hence, for proving \emph{(1)}, we can w.l.o.g. restrict ourselves to maps $T$ from the interior of $\cC_d$ or $\cL_d$, where we can furthermore exclude the set of channels with degenerate spectrum by Lem.\ref{lem:deg}. In other words, by subtracting a null set, we can begin with what is assumed in \emph{(1.b)} and \emph{(2.b)}, namely that $T$ has $d^2$ distinct eigenvalues. 
     If we denote the eigenvalues by $(\lambda_j)_{j=1}^{d^2}$, the Jordan decomposition of $T$ can be expressed as
     \begin{equation}
         \label{eq:JordanT}
         \tr{\rho T(H)}=\sum_{j=1}^{d^2} \lambda_j \tr{\rho Y_j} \tr{X_j^* H},
     \end{equation}
     with a biorthogonal system $\tr{X_j^* Y_k}=\delta_{j,k}$ for which we can assume that $Y_{d^2}=\1$ and $\lambda_{d^2}=1$ as $T$ is supposed to be unital.\vspace*{5pt}

    \emph{(1)} Under the above assumptions on the spectrum, we can exploit Thm.\ref{thm:seriesextension2} to extend the time series and in this way restrict ourselves to the case $t_0=0$, $t_1=d^2-2$. Injectivity of $\alpha$ is then equivalent to injectivity of $\alpha_0:\cH_d^{(0)}\rightarrow\R^{d^2-1}$, $\alpha_0(\sigma):=\big(\tr{\sigma T_Q^i(H_Q)}\big)_{i=0}^{d^2-2}$, where $T_Q:\cH_d^{(0)}\rightarrow\cH_d^{(0)}$, $T_Q:=Q\circ T= Q\circ T\circ Q$ and $H_Q:=Q(H_0)$. Using Eq.(\ref{eq:JordanT}) together with $\tr{\sigma Y_{d^2}}=0$ we obtain 
     $\alpha_0(\sigma)_i=\sum_{j=1}^{d^2-1} \lambda_j^i \tr{\sigma Y_j}\tr{X_j^* H_0}$. The transformation matrix $\hat{\alpha}_0\in\C^{(d^2-1)\times (d^2-1)}$ of the map $\alpha_0$ is therefore given by a matrix product $\hat{\alpha}_0=\Lambda M$, with $\Lambda,M\in\C^{(d^2-1)\times (d^2-1)}$, where $M_{jk}:=\delta_{jk}\tr{X_j^* H_Q}$ represents a diagonal matrix and $\Lambda_{ij}:=\lambda_j^i$ a \emph{Vandermonde matrix}. Since the determinant of the latter is known (\cite{horn13}, p.37) to be $\det(\Lambda)=\prod_{1\leq j<i\leq d^2-1} (\lambda_i-\lambda_j)\neq 0$, the matrix $\hat{\alpha}_0$ is invertible iff $\tr{X_j^* H_Q}\neq 0$ for all $j<d^2$. From here on, we  will distinguish the cases \emph{(1.a)} and \emph{(1.b)} of the theorem:

     \emph{(1.a)} For $H_0$ fixed, the equation $\det(\hat{\alpha}_0)=0$ can be expressed as a polynomial equation in the entries of $\hat{T}$. Therefore it is either fulfilled by a null set in $\cT_d$ or for the entire space. In order to exclude the latter, let $(X_j)_{j=1}^{d^2}$ be a basis of hermitian operators s.t. $\tr{X_j H_Q}\neq 0$ holds for all $j<d^2$, and for which  $(Y_k)_{k=1}^{d^2}$ with $Y_{d^2}=\1$ is the dual basis (w.r.t. the Hilbert-Schmidt inner product). Then we can define a  map $T:\cH_d\rightarrow\cH_d$ via $\tr{\rho T(H)}:=\sum_j \lambda_j \tr{\rho Y_i} \tr{X_j H}$ with $d^2$ distinct real eigenvalues $\lambda_j$. If $\lambda_{d^2}=1$, then $T\in \cT_d$ and we have $T\in\cC_d$ if all other eigenvalues are sufficiently small as $T$ then approaches the completely depolarizing channel, which lies in the interior of $\cC_d$. By construction, $T$ then gives rise to $\det(\hat{\alpha}_0)\neq 0$.

     \emph{(1.b)} For $j<d^2$, we have $\tr{X_j^* H_Q}=\tr{X_j^* H_0}$ so that  $O_j:=\{H_0\in\cH_d|\tr{X_j^* H_Q}=0\}$ is a proper subspace of $\cH_d$ and therefore a null set. Then, $\cup_{j=1}^{d^2-1}O_j$ is a null set as well.\vspace*{5pt}
     
     \emph{(2)} The proof is analogous to the proof of \emph{(1)} with the only main difference that we work with the entire $d^2$-dimensional space $\cH_d$ instead of a $(d^2-1)$-dimensional subspace. Using the Jordan decomposition of $T$ we obtain $\beta(H)_i=\sum_{i=1}^{d^2} \lambda_{j}^i \tr{\rho_0 Y_j}\tr{X_j^* H}$. The corresponding transformation matrix is then $\hat{\beta}=\Lambda M$, where $\Lambda$ is again an invertible Vandermonde matrix and $M_{ij}:=\delta_{ij}\tr{\rho_0 Y_j}$ so that $\hat{\beta}$ is invertible iff $\tr{\rho_0 Y_j}\neq 0$ for all $j$. \emph{(2.a)} then follows by copying the argument from the proof of \emph{(1.a)}. Similarly, \emph{(2.b)} follows by noting that $\{\rho_0\in\cH_d^{(1)}|\tr{\rho_0 Y_j}=0\}$ is a proper affine subspace of $\cH_d^{(1)}$ (or empty for $j=d^2$) and therefore has $(d^2-1)$-dimensional Lebesgue measure zero.
\end{proof}

The following is a continuous-time version of the foregoing theorem:
\begin{corollary}\label{cor:contmain2}
    Let $t_1,\ldots,t_{d^2}\in\R$ be distinct, $H_0\in\cH_d$ not proportional to the identity, and $\rho_0\in\cH_d\setminus\cH_d^{(0)}$. For every $L\in\cG_d$ except a null set, the following maps are injective:\margintxt{The first map requires one point less, but proving this would require a strengthening of Thm.\ref{thm:Lextension}.}
    \begin{eqnarray}
        \cH_d^{(1)}\ni\rho &\mapsto & \Big(\tr{\rho e^{t_k L}(H_0)}\Big)_{k=1}^{d^2}\in\R^{d^2},\nonumber\\
         \cH_d\ni H &\mapsto & \Big(\tr{\rho_0 e^{t_k L}(H)}\Big)_{k=1}^{d^2}\in\R^{d^2},\nonumber
    \end{eqnarray}
\end{corollary}
    \begin{proof}
        We first use Thm.\ref{thm:Lextension}, which enables an extension from the finite set of discrete times $t_1,\ldots,t_{d^2}$ to the entire real line. As this includes in particular all times $t\in\{1,\ldots,d^2\}$, we can directly employ Thm.\ref{thm:main1d2} \emph{(1.a)}, \emph{(2.a)}.
    \end{proof}

The next result is an analog of Takens' delay embedding theorem in the context of quantum state tomography. From a physical perspective, we assume that prior information restricts the set of states to a set of reduced (Minkowski) dimension. The theorem shows how this also reduces the number of data points that are sufficient for time-series-based quantum state tomography.

\begin{theorem}\label{thm:TakensQ}\margintxt{This could be formulated for observables as well.} Let $S\subset\cH_d$ be a closed subset of density operators and $H_0\in\cH_d$ not proportional to the identity. If $m>D(S-S)$, then for almost every quantum channel $T\in\cC_d$ the map
\begin{equation}\label{eq:Takensmapthm}
    S\ni\rho\ \mapsto\ \big(\tr{\rho T^i(H_0)}\big)_{i=1}^m
\end{equation} is injective on $S$ with a H\"older continuous inverse. More precisely, for any $\theta\in(0,1-D(S-S)/m)$ the subset of $\cC_d$ that gives rise to an injective map with $\theta$-H\"older continuous\footnote{A function $f$ is $\theta$-H\"older continuous if there is constant $c$ s.t. for all $x,y$: $\|f(x)-f(y)\|\leq c\|x-y\|^\theta$.} inverse has non-zero measure in $\cT_d$ and its complement is a null set. Moreover, the same is true when we replace $\cC_d$ by $\cL_d$.
\end{theorem}

\begin{proof} 
We can restrict ourselves to $m\leq d^2-1$, since the case $m\geq d^2$ is already covered by Thm. \ref{thm:main1d2}.

The basic idea of the proof is to reduce the considered scenario of $m$ highly dependent measurements, which appear in Eq.(\ref{eq:Takensmapthm}), to the case of $m$ independent measurements. To this end, we divide the proof into three steps: 1. We define the map $\tau:\cT_d\rightarrow\cH_d^m$, $\tau(T):=\big(T^k(H_0)\big)_{k=1}^m$ and show that $\tau(\cL_d)$ has non-zero measure. 2. We prove that $\tau$ has the Luzin $N^{-1}$ property, i.e.,  that preimages of null sets are null sets. 3. We exploit the previous steps in order to apply a known theorem, which states that $m$ independent measurements are generically sufficient and give rise to a H\"older continuous inverse.\vspace*{5pt}

    \emph{Step 1.}: Consider any $T$ in the interior of $\cL_d$ and such that  $\1,H_0,T(H_0),\ldots,$ $T^{d^2-2}(H_0)$ form a set of $d^2$ linearly independent operators. By Lemma \ref{lem:linindep} this is true for almost every $T\in\cL_d$. Define $\epsilon:\cH_d^m\rightarrow[0,\infty)$ as
    \begin{equation}
        \epsilon(A):=\sup\big\{\delta\in\R\;|\;\forall\varepsilon\in[0,\delta]\;\exists T_\varepsilon\in\cL_d:\tau(T_\varepsilon)=\tau(T)+\varepsilon A\big\}.
    \end{equation}
We want to show that $\epsilon(A)>0$ for every $A$. In order to construct the desired $T_\varepsilon$, we introduce four auxiliary maps $\Phi_i,\Psi_i:\cH_d\rightarrow\cH_d$ for $i\in\{0,1\}$ as follows: let $E_1,\ldots,E_{d^2}$ be a basis of $\cH_d$ with $E_1=\1$ and impose that $\Phi_i:E_1\mapsto\1$, $\Psi_i:E_1\mapsto 0$ and for $k\in\{0,d^2-2\}$:
\begin{eqnarray}
\Phi_i:E_{k+2} &\mapsto&  T^{k+i}(H_0)\\
\Psi_i: E_{k+2}&\mapsto&  A_{k+i} \quad \text{with }A_j:=0 \text{ if }j\not\in\{1,\ldots,m\}.
\end{eqnarray}
Note that such maps always exist (since any set of vectors can be reached by a linear map acting on any given basis). Moreover,  the assumed linear independence guarantees that $\Phi_0$ is invertible as  it maps one basis onto another one. Hence, for sufficiently small $\varepsilon>0$ we can define $T_\varepsilon\in\cT_d$
\begin{eqnarray}
T_\varepsilon&:=&\big(\Phi_1+\varepsilon\Psi_1\big)\big(\Phi_0+\varepsilon\Psi_0\big)^{-1},\quad \text{which maps }\\
   T^k(H_0)+\varepsilon A_k  &\mapsto & T^{k+1}(H_0)+\varepsilon A_{k+1},
\end{eqnarray}
   for all $k\in\{0,d^2-2\}$. Consequently, $\tau(T_\epsilon)=\tau(T)+\varepsilon A$ and since $\lim_{\varepsilon\rightarrow 0}T_\varepsilon=T$ and $T$ was chosen from the interior of $\cL_d$, we have $\epsilon(A)>0$ indeed. This proves that for every non-empty open subset $U\subset\cL_d$ the set $\tau(U)$ has non-zero measure.
 \vspace*{5pt}

    \emph{Step 2.}: We begin by showing that the set of critical points of $\tau$, i.e., the set of $T$'s for which the differential ${\mathrm d}\tau(T)$ is not surjective, is a null set. Since $\tau$ is a polynomial map, the use of  minors shows that the set $C\subset\cT_d$ of critical points  can be expressed as  the zero set of a polynomial. As such, this set is either a null set or the entire space (if the polynomial itself is zero). Suppose it were the entire space. Then its image under $\tau$, the set of critical values, would have non-zero measure according to \emph{Step 1}. However, this contradicts Sard's theorem (cf. \cite{Milnor1965}, p.10), according to which the set of critical values, and in our case therefore also the set of critical points, is a null set. 

    By the constant rank theorem (cf. \cite{tu2010introduction}, Thm.11.1), every regular point $x\in\cT_d\setminus C$ has an open neighborhood $U$ in which $\tau|_U$ is diffeomorphic to a coordinate projection and therefore such that preimages of null sets are null sets. 

So if $N\subset \cH_d^m$ is an arbitrary null set, then there is a countable open cover $\bigcup_n U_n\supseteq \cT_d\setminus C$ such that
\begin{equation}
    \tau^{-1}(N)\subseteq C\cup\bigcup_n \big(\tau|_{U_n}\big)^{-1}(N)
\end{equation} is contained in a countable union of null sets and  thus a null set itself.
 \vspace*{5pt}

    \emph{Step 3.}: Now we are positioned to resort to a well-known theorem, which states that for $m>D(S-S)$ the subset in $\cH_d^m$ that does not lead to a $\theta$-H\"older continuous inverse is a null set (see \cite{robinson_2010}, Thm.4.3 and \cite{Heinosaari2013} for its application in quantum tomography). If we denote this set by $N\subset \cH_d^m$, then we know from \emph{Step 2.} that $\tau^{-1}(N)$ is a null set in $\cT_d$, which completes the proof.
    
\end{proof}
The following Lemma has been used in the proof:
\begin{lemma}\label{lem:linindep} If $m\leq d^2-1$ and $H_0\in\cH_d$ is not proportional to the identity, then $S_0:=\{T \in \cT_d\;|\; \1,H_0,T(H_0),\ldots,T^{m-1}(H_0) \text{ lin. dependent} \}$ is a null set.
\end{lemma}
\begin{proof} We build a matrix $M\in \R^{d^2\times (m+1)}$, whose columns are given by the images of $\1,H_0, T(H_0),\ldots,T^{m-1}(H_0)$ under $\nu$. Then $T\in S_0$ iff all $(m+1)\times (m+1)$ minors of $M$ vanish. That is, there is a real polynomial function $p$ in the entries of the transformation matrix $\hat{T}$ s.t. $T\in S_0\Leftrightarrow p(\hat{T})=0$. Hence, $S_0$ is a null set unless $p$ is constant zero. In order to exclude the latter, it suffices to provide a single example $T\in\cT_d\setminus S_0$. To construct such an example, we exploit the fact that there always exists a linear map that maps a given basis to a given set of vectors. Specifically, if we let $H_0,\ldots,H_{d^2-1}$ with $H_{d^2-1}:=\1$ be any basis of $\cH_d$, there is a $T\in\cT_d$ s.t. $T:\1\mapsto\1$, $T:H_{d^2-2}\mapsto H_0$ and $H_{i-1}\mapsto H_{i}=T^i(H_0)$ for all $i\in\{1,d^2-2\}$. Since $T$ simply permutes basis elements, $H_0,T(H_0),\ldots,T^{m-1}(H_0)$ are  linearly independent by construction.
\end{proof}

\section{Stability and finite statistics}\label{sec:stability}

In this section, we will analyze statistical aspects, which become relevant if probabilities or expectation values are not known exactly but only estimated empirically on the basis of a finite number of measurement outcomes. Our aim is to prove bounds on the quality of statistical estimators in terms of the smallest singular values of the maps $\alpha$ and $\beta$ that appear in Thm.\ref{thm:main1d2}. Along the lines of \cite{Guta_2020} and \cite{confidenceregions}, we consider a least squares estimator.

Consider  $\alpha(\rho):=\big(\tr{\rho H_i}\big)_{i=0}^{d^2-2}$ and assume that each  $H_i:=T^i(H_0)$ is measured $n$ times, independently. That is, in total $n(d^2-1)$ independent measurements are performed on $n(d^2-1)$ independent and identically prepared systems, each described by the density operator $\rho$. Let  $f_i$ be the empirical estimate for $\tr{H_i\rho}$, obtained by averaging the outcomes of the $n$ individual measurements. Abusing notation and denoting by $f_i$ also the corresponding random variable,  $\mathbb{E}[f_i]=\tr{H_i\rho}$ is the corresponding expectation value. Moreover, if $H_i$ is the effect operator of a  POVM, $\tr{H_i\rho}$  is the probability of obtaining the corresponding outcome, so that $f_i$ follows a binomial distribution with variance ${\rm Var}[f_i]=\tfrac{1}{n}\tr{H_i\rho}(1-\tr{H_i\rho})\leq\tfrac{1}{4n}$.

If $\alpha$ is injective as a map from the set of density operators, then, due to linearity, $\alpha(\rho_1-\rho_2)=0$ holds for two density operators $\rho_1, \rho_2$ only if $\rho_1-\rho_2=0$. In other words, $\alpha$ is then invertible as a linear map from the space of trace-less Hermitian matrices $\cH_d^{(0)}$ into $\R^{d^2-1}$. In the following, the inverse $\alpha^{-1}$ will always be understood from this perspective, i.e.,  $\alpha^{-1}:\R^{d^2-1}\rightarrow\cH_d^{(0)}$.

Consider an estimator $\hat{\rho}$ that is a minimizer of \begin{equation}
    \min_X \Big\{\underbrace{\sum_{i=0}^{d^2-2} \left(\tr{H_i X}-f_i\right)^2}_{=\|\alpha(X)-f\|_2^2}\big|\; X=X^*\wedge \tr{X}=1\Big\}.\label{eq:rhohut103}
\end{equation}
More explicitly, we choose $\hat{\rho}:=\rho_0+\alpha^{-1}\big(f-\alpha(\rho_0)\big)$, where $\rho_0$ is an arbitrary reference density matrix, e.g. $\rho_0:=\1/d$. This choice is hermitian, has unit trace and satisfies $\|\alpha(\hat{\rho})-f\|_2=0$ therefore minimizing Eq.(\ref{eq:rhohut103}). 

\begin{theorem}\label{thm:errorbound} In the above scenario, the distance between the true density matrix $\rho$ and the estimator $\hat{\rho}$, which is computed from the outcomes of $n(d^2-1)$ independent measurements, satisfies:  
\begin{equation}\label{eq:iug39j}
    \mathbb{E}\Big[\|\hat{\rho}-\rho\|_2^2\Big]\ \leq\  \|\alpha^{-1}\|^2\sum_{i=0}^{d^2-2} {\rm Var}[f_i]\  \leq \ \frac{\|\alpha^{-1}\|^2\Delta^2 (d^2-1)}{4n},
\end{equation}
where $\Delta\in\R_+$ is such that the spectrum of every $H_i$ is contained in an interval of length at most $\Delta$. $\|\alpha^{-1}\|$ denotes the operator norm of $\alpha^{-1}$ when regarded as a linear map from $\R^{d^2-1}$ into the space of traceless hermitian matrices, equipped with the Euclidean and Hilbert-Schmidt norm, respectively.
\end{theorem}

\begin{proof}
    Inserting $\hat{\rho}$ and using the linearity of the expectation value we obtain
    \begin{eqnarray}
        \mathbb{E}\Big[\|\hat{\rho}-\rho\|_2^2\Big] &=& \mathbb{E}\Big[\|\rho_0+\alpha^{-1}\big(f-\alpha(\rho_0)\big)-\rho\|_2^2\Big]\nonumber\\&=& \nonumber\mathbb{E}\Big[\|\alpha^{-1}\big(f-\alpha(\rho_0)\big)-\alpha^{-1}\alpha(\rho-\rho_0)\|_2^2\Big]\\&\leq& \nonumber\|\alpha^{-1}\|^2 \mathbb{E}\Big[\|f-\alpha(\rho)\|_2^2\Big]\\ &=& \nonumber\|\alpha^{-1}\|^2 \sum_{i=0}^{d^2-2} \underbrace{\mathbb{E}\Big[\big(f_i-\tr{H_i\rho}\big)^2\Big]}_{={\rm Var}[f_i]}.
    \end{eqnarray}
    Since the variance does not change under translation, we can w.l.o.g. assume that the range of the outcomes of the $i$'th observable is $[-\tfrac{\Delta}{2},\tfrac{\Delta}{2}]$. The maximal variance under this constraint is $\tfrac{\Delta^2}{4}$, which is reduced by a factor $\tfrac{1}{n}$ when averaging $n$ independent outcomes.
\end{proof}

In order to get a reference point for comparison and later use, let us for a paragraph depart from the framework of evolution-based tomography and consider the measurement of independent observables. Suppose these are described by a set of Hilbert-Schmidt orthogonal Hermitian operators $H_i$ that satisfy $\tr{H_i H_j}=\delta_{ij}d$ (as it is, for instance, the case for Pauli matrices or tensor products thereof). In this case, the corresponding map $\alpha$ satisfies $\alpha\alpha^*=d \1$ so that $\|\alpha^{-1}\|^2=1/d$.\vspace*{5pt}

For a map $\beta:\cH_d\rightarrow\R^{d^2}$, $\beta(H):=\big(\tr{\rho_i H}\big)_{i=0}^{d^2-1}$ with $\rho_i:=T^{* i}(\rho_0)$ we can now argue similarly. Suppose $H$ is an effect operator of a binary POVM, i.e., that $\tr{\rho_i H}$ is the probability of obtaining the corresponding measurement outcome if the preparation is described by $\rho_i$. Denoting the relative frequencies again by $f_i$, so that $\mathbb{E}[f_i]=\tr{\rho_i H}$ and using the estimator $\hat{H}:=\beta^{-1}(f)$, the exact same reasoning as in the proof of Thm.\ref{thm:errorbound} leads to
\begin{equation}
    \mathbb{E}\big[\|H-\hat{H}\|_2^2\big]\leq \frac{\|\beta^{-1}\|^2 d^2}{4n}.
\end{equation}

Let us again compute a reference point for $\|\beta^{-1}\|^2$ in which we detach the $\rho_i$'s from the requirement that they have to be related by a time-evolution. Specifically, we consider a set of pure states $\rho_1,\ldots,\rho_{d^2}$ that form a SIC-POVM. Then $\beta\beta^*$ is is represented by a $d^2\times d^2$ Gram-matrix with entries $$\tr{\rho_i\rho_j}=\frac{1+\delta_{ij}d}{1+d}.$$
This matrix has spectrum $\{d,d/(d+1)\}$. Consequently, $\|\beta^{-1}\|^2=(d+1)/d$ in this case.\margintxt{One might try to prove that this is indeed the optimal value.}\vspace*{10pt}

\section{The curious case of qubits}\label{sec:qubits}
The case of qubits is special in several ways. In particular, there is a drastic difference between time-series-based tomography of states and that of observables since the no-go results of Cor.\ref{cor:nogo2} apply to the latter but not to the former when applied to unitary evolution:\vspace*{5pt}

{\bf Tomography of states:} In this case, Cor.\ref{cor:nogo} and Cor.\ref{cor:nogo2} do not imply a no-go result since  $D(\cH_d^{(1)})=d^2-1$ equals $d^2-d+1$ for $d=2$.
In fact, full state tomography based on the time series of a unitary qubit channel is not only possible, it turns out to be possible in an optimal way---even when compared to state tomography that is not bound to a single observable undergoing a homogeneous time evolution. This is a consequence of another qubit-specialty, namely that $SU(2)/\{\pm\1\}\simeq SO(3)$. This enables the existence of a unitary channel $T\in\cC_d$, $T(H)=U^* H U$ which implements a cyclic permutation of basis elements. More precisely, if $\sigma_1,\sigma_2,\sigma_3$ are the three Pauli matrices, we have $\sigma_1\stackrel{T}{\mapsto}\sigma_2\stackrel{T}{\mapsto}\sigma_3\stackrel{T}{\mapsto}\sigma_1$ for
\begin{equation}
    U=\tfrac12\begin{pmatrix}
        i+1 & 1+i\\ i-1 & 1-i
    \end{pmatrix} \ =\ \exp\left[\tfrac{i}{2}\theta \sum_{k=1}^3 w_k\sigma_k\right],
\end{equation}
where $\theta:=2\pi/3$ is the rotation angle and $w:=(1,1,1)/\sqrt{3}$ the rotation axis. Hence, with $H_0=\sigma_3$ we get 
\begin{equation}
    \left(\tr{\rho T^i(H_0)}\right)_{i=1}^3\ = \ \big(\tr{\rho\sigma_i}\big)_{i=1}^3 ,
\end{equation} which is not only informationally complete but optimal in almost every conceivable way.\vspace*{5pt}

\begin{figure}[t]
    \centering
    \includegraphics[width=0.78\textwidth]{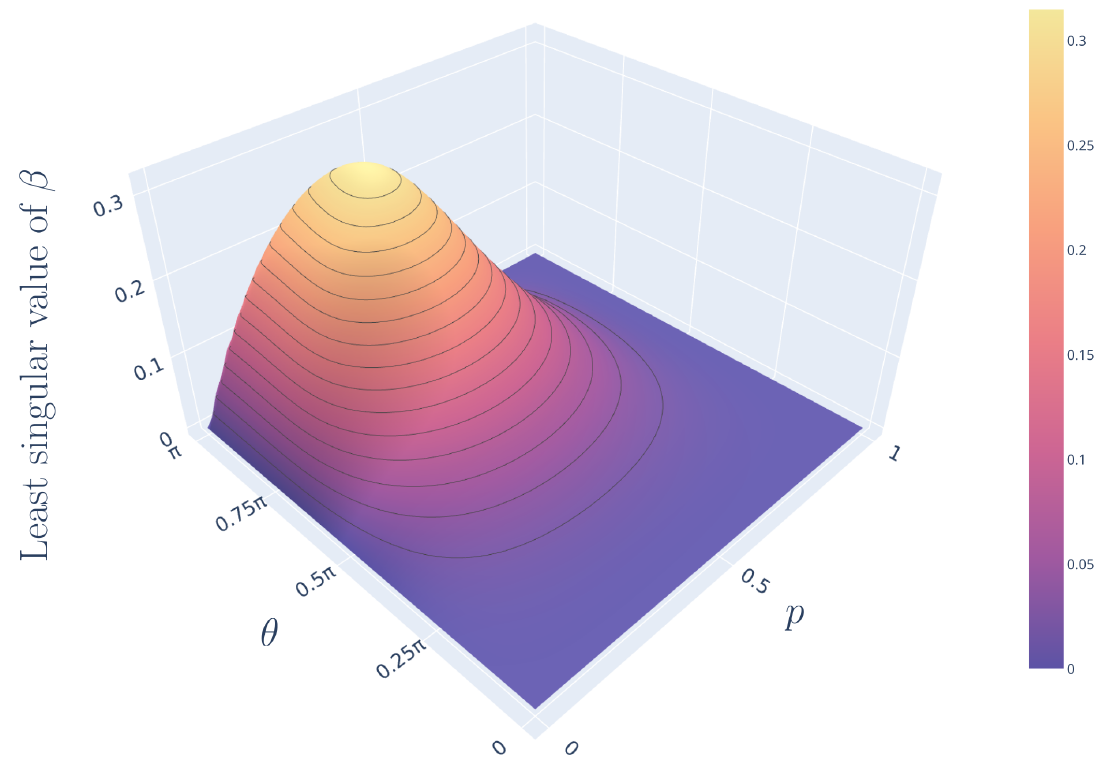} 
    \caption{The least singular value of the map $\beta$ from Eq.(\ref{eq:qubitbeta}) for the qubit-channel of Eq.(\ref{eq:qubitchannel}) quantifies how stably the map can be inverted, and (according to Sec.\ref{sec:stability}) how much statistics is required. Here, $p$ quantifies the amount of depolarizing noise, and $\theta$ the degree of unitary rotation. While a unitary evolution ($p=0$) prohibits invertibility, additional noise ($p>0$) enables invertibility and thus full tomography of qubit observables.\vspace*{-0pt}} \label{fig:3}
\end{figure}

{\bf Tomography of observables:}  For the case of observables, Cor.\ref{cor:nogo} and Cor.\ref{cor:nogo2} imply that full tomography is not possible in a time-series-based manner for any unitary evolution in any dimension $d\geq 2$ since 
 $D(\cH_d)=d^2>d^2-d+1\geq \delta(T)$. Furthermore, for $d\geq 3$ Cor.\ref{cor:nogo2}  shows that adding simply depolarizing noise does not help. However, for $d=2$ simply depolarizing noise is just enough to enable full tomography. To illustrate this situation, consider the qubit channel 
 \begin{equation}\label{eq:qubitchannel}
     T(H):=(1-p)e^{-i\tfrac{\theta}{2}\sigma_z} H e^{i\tfrac{\theta}{2}\sigma_z}+p \1\tr{H}/2,
 \end{equation}
  depending on a parameter $p\in[0,1]$, which quantifies the amount of depolarizing noise, and $\theta\in[0,\pi]$,  quantifying the degree of rotation. 
 Suppose the initial state is pure and an equal weight superposition of the $+1$ eigenstates of the three Pauli matrices, i.e., $|\psi\rangle\propto|+_x\rangle+|+_y\rangle+|+_z\rangle\propto(2+\sqrt{2})|0\rangle+(1+i)|1\rangle$.
 
 The map
 \begin{equation}\label{eq:qubitbeta}
     \cH_2\ni H\mapsto\beta(H):=\big(\langle\psi|T^i(H)|\psi\rangle\big)_{i=0}^3
 \end{equation} 
 is injective if its smallest singular value is non-zero. As discussed in Sec.\ref{sec:stability}, the magnitude of this singular value is related to the stability of the inversion and to the amount of required statistics. The smallest singular value is readily computed numerically and displayed in Fig.(\ref{fig:3}). This shows that, starting from a non-trivial unitary rotation, $\beta$ becomes injective as soon as depolarizing noise is added. As expected, if too much noise is added, the stability of the inversion decreases and the map becomes non-invertible again for $p=1$.   

Finally, let us remark that there are qubit channels, with extremal cases lying on the boundary of the set of channels, for which a significantly higher least singular value of $\beta$ (up to $0.475$) can be achieved numerically. Unfortunately, it appears difficult to derive an analytic expression for the optimum.

\section{Open questions}

The focus of this work is on answering the fundamental qualitative question of when quantum tomography is feasible in principle within the specified evolution-based framework. However, many of the more quantitative questions remain open for now. For instance: How should limited statistics be distributed over the different measurements? How should time intervals be chosen?  What are the optimal channels in general and, for a given Hamiltonian evolution, what is the optimal noise that should be added for optimal evolution-based tomography? 
\vspace*{5pt}\\ 


\emph{Acknowledgments:} MMW acknowledges funding by the Deutsche Forschungsgemeinschaft (DFG, German Research Foundation) under Germany's Excellence Strategy –  EXC-2111 – 390814868 and via the TRR 352 – Project-ID 470903074. HR acknowledges funding from the Munich Quantum Valley. The authors also acknowledge interesting discussions with Alvaro Yanguez Bachiller, whose Master's thesis \cite{Alvaro} was based on the ideas of the present paper and contains an investigation of further numerical examples.

We thank Christian Arenz, Steven Flammia, Francesco Ticozzi, and Frederik vom Ende for bringing relevant literature to our attention.

\bibliographystyle{halpha}
\bibliography{QuantumTakens}{}

\end{document}